# Explainable Risk Classification in Financial Reports

Xue Wen Tan
*National University of Singapore*, xuewen@u.nus.edu

Stanley Kok
*National University of Singapore*, skok@comp.nus.edu.sg

# Explainable Risk Classification in Financial Reports

*Completed Research Paper*

**Xue Wen Tan**
Asian Institute of Digital Finance
National University of Singapore
xuewen@u.nus.edu

**Stanley Kok**
Department of Information Systems and Analytics
National University of Singapore
skok@comp.nus.edu.sg

## Abstract

*Every publicly traded company in the US is required to file an annual 10-K financial report, which contains a wealth of information about the company. In this paper, we propose an explainable deep-learning model, called FinBERT-XRC, that takes a 10-K report as input, and automatically assesses the post-event return volatility risk of its associated company. In contrast to previous systems, our proposed model simultaneously offers explanations of its classification decision at three different levels: the word, sentence, and corpus levels. By doing so, our model provides a comprehensive interpretation of its prediction to end users. This is particularly important in financial domains, where the transparency and accountability of algorithmic predictions play a vital role in their application to decision-making processes. Aside from its novel interpretability, our model surpasses the state of the art in predictive accuracy in experiments on a large real-world dataset of 10-K reports spanning six years.*

## Introduction

The financial sector has long relied on traditional risk classification methods, such as analyzing credit scores, scrutinizing debt-to-income ratios, and evaluating financial statements (Altman, 1968). While these approaches have been effective in certain risk-related applications, like assessing the riskiness of borrowers, investments, and financial instruments, they are inherently limited in scope and necessitate time-consuming human intervention. To ameliorate these limitations, recent literature on financial risk analysis has predominantly focused on statistical methods (David et al., 2011; Fama & French, 1993; Toma & Dedu, 2014). In a notable study (Fama & French, 1993), such techniques successfully identified a firm's size and its book-to-market ratio as primary factors influencing financial risk, but fell short in revealing additional contributing factors. This underscores the limitations of these methods in uncovering hidden patterns and complex interactions, particularly within the vast and complex datasets prevalent in today's modern financial environment.

In the realm of financial risk classification, a powerful tool has emerged in the form of machine learning. This tool improves risk assessment by effectively utilizing large datasets and advanced algorithms, producing models that are adept at predicting potential risk factors with precision and uncovering complex relationships between variables (Shi et al., 2022). This technological advancement has substantially enhanced risk assessment capabilities, allowing financial organizations to make well-informed decisions in a fraction of the time required by traditional methods or basic statistical techniques.

The financial industry has also experienced significant transformation due to advancements in a subfield of machine learning called natural language processing (NLP). This technique allows for the extraction of valuable insights from unstructured textual data, which can then be applied to improve risk evaluation and investment decision-making. The success of NLP has led researchers to explore alternative textual data sources, such as news articles, reviews, and financial reports, in search of additional risk factors. Simultaneously, researchers have investigated various text analytics techniques to extract further valuable insights (Ding et al., 2015; Kogan et al., 2009; Nopp & Hanbury, 2015; Rekabsaz et al., 2017; Tsai & Wang, 2017). Foremost among these techniques is the method of deep learning which stands out due to its ability to learn semantically meaningful representations without the need for extensive manual feature engineering. Through its different manifestations, such as CNN (Lecun et al., 1998), GRU (Chung et al., 2014), and BERT (Devlin et al., 2018), deep learning has demonstrated impressive results in various NLP tasks like document classification and sentiment analysis (Akhtar et al., 2017; Dos Santos & Gatti, 2014).

However, deep learning comes with its own set of drawbacks. One of the most notable is its black-box nature, which has raised concerns among regulators, stakeholders, and financial institutions (Authority, 2018). The opacity of deep learning models complicates the understanding of how they reach specific decisions or predictions. This lack of transparency can result in distrust toward deep learning systems, particularly when significant decisions affecting individuals or organizations are based on the output of these models (Ribeiro et al., 2016). In some instances, organizations may opt to avoid using them altogether, even if they exhibit strong performance (Arnold et al., 2006; Kayande et al., 2009). Moreover, this explainability deficit can impede compliance with regulatory requirements, as financial institutions are frequently obligated to justify their risk assessment processes (Bussmann et al., 2021).

Because of these concerns, there is a growing need in financial risk classification for artificial intelligence (AI) models that can provide clear and understandable reasons for their decisions, commonly referred to as explainable AI. By combining the power of artificial intelligence with explainability, explainable AI can create a more transparent and accountable system, which is essential for maintaining trust and compliance in the financial sector (Došilović et al., 2018). When incorporated into financial risk classification models, explainable AI can foster a deeper understanding of risk-contributing factors, empowering financial institutions to make well-informed decisions regarding lending, investments, and risk management (Bussmann et al., 2021). Additionally, explainable AI can help uncover previously overlooked factors related to risk and detect potential biases in models, resulting in more accurate and fair risk assessments (Linardatos et al., 2020; Meske et al., 2022).

The majority of existing explainable AI approaches in NLP primarily focuses on providing interpretability at the *word level* (Sun & Lu, 2020; Tjoa & Guan, 2020; Zhao et al., 2020) or occasionally at the *sentence level* (Lin et al., 2021; Mathews, 2019). While these explanations provide some insight into the models' inner workings, they often fail to provide a comprehensive, high-level explanation spanning an entire large corpus. There are several critical reasons for providing high-level explanations at the *corpus level*. First, it enables users to grasp the broader context of risk factors, ensuring they do not miss the forest for the trees. By identifying overarching themes and trends in the entire corpus, users can recognize wide-ranging patterns that might be overlooked when focusing solely on word-level or sentence-level explanations. Second, high-level explanations promote effective communication between AI-driven models and their users. Financial professionals, stakeholders, and regulators often need to interpret and explain complex risk assessments to others, and these high-level explanations allow them to present a coherent and comprehensive narrative that connects all significant risk factors. Third, high-level explanations empower users to make more informed decisions in their risk management practices. By identifying extensive areas of concern, these explanations allow for well-coordinated financial interventions, leading to a more robust approach of managing financial risk.

To further advance the explainable AI field in the finance sector, we introduce a model called the FinBERT eXplainable Risk Classifier (FinBERT-XRC). Our FinBERT-XRC model addresses the limitations of existing explainable AI models by providing a comprehensive, 360-degree explanation of the factors that contribute to a company's risk classification. FinBERT-XRC is specifically designed to read and analyze 10-K documents, which are essential for understanding a company's financial health. FinBERT-XRC leverages the power of NLP methods and explainable AI techniques to accurately determine a company's risk profile while providing a multi-layered explanation of its decision-making process.

Our FinBERT-XRC model distinguishes itself from existing models by providing explanations simultaneously at *all* three of the aforementioned levels. The first level is the word level, where the model highlights the importance of specific words within a sentence, allowing users to understand which words contribute the most to the overall risk assessment. The second level is the sentence level, where the model highlights the most relevant sentences within a document. By identifying these critical sentences, users can focus on the most pertinent information and gain a better understanding of the factors affecting the company's risk profile. The third level, unique to FinBERT-XRC, is the corpus level. Here, the model presents word clouds that display the most relevant words, in an *entire corpus*, that are associated with the risk elements for a given year. To generate these word clouds, a specialized filtering layer is applied to our FinBERT-XRC model. This filter is based on the importance of each word and each sentence as determined by FinBERT-XRC's deep-learning algorithm, ensuring that only the most relevant words are counted. As a result, the word clouds effectively explain the risk classification and help users identify overarching themes and trends within the corpus, providing invaluable context for understanding the broader risk landscape.

By offering explanations at multiple levels, our FinBERT-XRC model addresses the twin challenges of transparency and efficacy in AI-driven financial risk classification. Our comprehensive approach not only enables financial professionals and stakeholders to trust the model's output but also empowers them to make better-informed decisions in their risk management practices. In essence, FinBERT-XRC serves as a powerful solution that combines the power of AI-driven analysis with the transparency and interpretability required in the financial sector, paving the way for effective and responsible financial risk assessment.

In summary, our FinBERT-XRC model advances the state of the art in the following key areas.

1. **Multi-level explanations.** FinBERT-XRC introduces a unique three-level explanation framework, which includes word, sentence, and corpus levels. This comprehensive approach allows users to gain a deeper understanding of the factors influencing a company's risk profile and provides valuable insights into the broader financial risk landscape.
2. **Enhanced classification performance.** In empirical comparisons on a real-world dataset, not only does FinBERT-XRC excel at providing explanations for its predictions, it also demonstrates superior classification performance in terms of F1 scores compared to existing strong baselines.

# Related Work

## *Text Analytics in the Financial Domain*

Bellstam et al. (2021) employed text analytics to assess the innovativeness of companies based on analyst reports for S&P 500 firms. Their text-based approach successfully forecasted firm performance and growth opportunities for up to four years. Similarly, Das et al. (2016) provided an overview of systems and methods for tracking ongoing events from sources such as corporate filings, financial articles, analyst reports, press releases, customer feedback, and news articles. Both studies utilize Latent Dirichlet Allocation (LDA) for classification tasks. LDA is a widely used statistical model in natural language processing (NLP) for topic modeling. It identifies underlying topics within a collection of text documents and estimates their relative frequency. LDA assumes that each document consists of a mixture of topics, with each topic represented by a distribution of words. By analyzing word co-occurrence in documents, LDA uncovers hidden thematic structures and reveals prevalent topics. However, LDA has a notable limitation: it disregards the order of words within the text, potentially leading to the loss of vital information. This issue arises from LDA's reliance on the bag-of-words representation, which only considers word presence and frequency in a document, but not their sequential arrangement. Consequently, LDA might struggle to capture contextual and syntactical nuances within texts, which are critical for understanding the true meaning and significance of the content, especially in complex domains like financial document analysis. Despite these limitations, LDA remains a popular choice in finance-specific applications due to its simplicity, interpretability, and ease of understanding, which are crucial in the finance domain. In contrast, transformer models like FinBERT (Liu et al., 2021), despite their superior performance, have yet to gain widespread adoption due to their “black box” nature.

The empirical effectiveness of bag-of-words techniques, such as LDA, has been surpassed by the deep-learning approach of word embeddings. Word embeddings represent words as numerical vectors that encapsulate both syntactic and semantic associations between words. They are learned from large collections of text data and can be used for various NLP tasks, such as text classification and sentiment analysis. Yang et al. (2022) built on recent progress in representation learning and proposed a novel word-embedding method that incorporates external knowledge from a finance-domain lexicon (Loughran &

McDonald, 2011). This approach learns semantic relationships among words in firm reports for better stock volatility prediction. Its empirical results demonstrate that domain-lexicon-enhanced text representation learning significantly outperforms bag-of-words models and generic word embeddings for stock volatility prediction.

The recent eXplainable Risk Ranking (XRR) model (Lin et al., 2021) is currently the state-of-the-art system for risk classification in financial reports. It features a bi-level system that provides explanations at the word and sentence levels. XRR trains domain-specific word embeddings using 10-K financial documents by fine-tuning pre-trained static word embeddings such as GLoVe (Pennington et al., 2014). Similar to Yang (2022), their word embeddings are also static, meaning that a word will have the same vector regardless of the context in which it is placed. Consequently, such embeddings cannot adequately represent polysemous words whose semantics depend on their contexts. In contrast, our FinBERT-XRC model uses *context-dependent* word embeddings, thereby outperforming XRR in terms of both interpretability and predictive accuracy. Furthermore, FinBERT-XRC simultaneously provides explanations at three levels -- the word, sentence and corpus levels, while XRR's explanations remain confined to the word and sentence levels only.

## *Post-hoc Explainability Techniques*

Two popular explainable AI frameworks that provide model-agnostic explanations for machine learning models are Local Interpretable Model-agnostic Explanations (LIME; Ribeiro et al., 2016) and SHapley Additive explanation (SHAP; Lundberg & Lee, 2017). LIME generates locally linear explanations for individual predictions, while SHAP uses cooperative game theory to determine the contribution of each feature to a prediction.

In the financial sector, these techniques have been applied to areas such as credit risk classification, in which machine learning models are used to predict the likelihood of borrowers defaulting on loans (Gramegna & Giudici, 2021). LIME and SHAP assist financial institutions in understanding the factors influencing the models' predictions to promote transparency and enable better decision-making.

However, these techniques exhibit limitations. LIME's explanations may not represent the model's global behavior and can be sensitive to the choice of neighborhood and local model complexities. Although SHAP provides more consistent explanations, it can be computationally expensive, particularly for high dimensional datasets such as text data. Additionally, both techniques face challenges when applied to text inputs, as they rely on perturbing input features to estimate local models or feature attributions. Since the meaning of text inputs is encoded in their sequential structure, this perturbative approach often results in the generation of non-representative, nonsensical text inputs. Despite these limitations, LIME and SHAP continue to offer valuable insights for *non-text-based* inputs.

Another powerful tool for providing post-hoc explanations is the attention mechanism of a deep learning system. By assigning varying importance weights to different input elements, the attention mechanism allows a deep-learning model to focus on the most relevant parts of the input when making predictions (Vaswani et al., 2017). These attention weights can be interpreted as feature importance scores, offering insights into which parts of the input contribute the most to a model's prediction. This interpretability aspect of the attention mechanism makes it an attractive choice for generating explanations (Wiegreffe & Pinter, 2019). Compared to LIME and SHAP, the attention mechanism offers several advantages. It inherently captures the relationships between the input elements in a sequential structure, which is particularly important for textual data. Additionally, attention-based explanations are directly derived from the model's internal parameters, eliminating the need for additional computation or approximation (like those needed by LIME and SHAP). This results in more efficient and accurate explanations that are better aligned with the model's predictions. For this reason, our FinBERT-XRC model innovates by building upon attention mechanisms.

In the field of Information Systems (IS), interpretability in AI is a pressing concern, as reflected by recent research. Someh et al. (2022) emphasize the need for businesses to develop AI explanation capabilities, highlighting four critical areas—decision tracing, bias remediation, boundary setting, and value formulation—to address challenges such as opacity and model drift. Our FinBERT-XRC model aligns with "decision tracing" by offering a clear pathway to understand how predictions are made. Zhang et al. (2020) examine the use of AI in legal practice, illustrating its potential to transform businesses while also shedding light on the unique challenges of developing machine-learning AI systems, especially in knowledge-intensive work. Additionally, Asatiani et al. (2020) present a case study at the Danish Business Authority, offering a framework and recommendations to deal with the complexities of explaining the behavior of black-box AI systems and to avoid any potential legal or ethical issues. Similarly, our paper extends the application of explainable AI into the finance domain. Together, these studies reveal different facets of a shared problem: making AI transparent and comprehensible. In line with these works, our FinBERT-XRC research contributes to the field by introducing an explainable AI model that not only classifies financial risk but also provides clear and understandable explanations for the prediction.

# Background

## *10-K Financial Reports*

The U.S. Securities and Exchange Commission (SEC) requires a publicly traded company to provide a comprehensive summary of its financial performance in the form of an annual document called the 10-K financial report. This report encompasses a broad range of information, including financial statements, management's discussion and analysis (MD&A), business descriptions, risk factors, corporate governance, and executive compensation. By providing this detailed and audited insight into a company's financial health, business strategy, and potential risks, the 10-K report serves as an invaluable resource for investors, analysts, and regulators.

Each section of the report provides its own unique and valuable insights. For instance, through the report's presentation of balance sheets, income statements, and cash flow statements, analysts can assess a company's profitability, liquidity, and overall financial stability. Furthermore, the MD&A section, which presents management's viewpoint on the company's performance and future prospects, enables investors to comprehend the strategic direction and management's plans for tackling challenges and seizing opportunities. In addition, the risk factors section, which details potential risks and uncertainties that the company faces, allows investors to gain insights into the possible challenges that could impact a company's performance.

Given the wealth of information contained in the 10-K reports, researchers have employed natural language processing (NLP) techniques to automatically extract valuable insights from the vast amounts of unstructured textual data found in these financial reports, uncovering patterns and connections that might not be easily discernible through manual analysis. The information and insights extracted can be used to facilitate numerous tasks such as sentiment analysis (Li, 2010; Ren et al., 2013), topic modeling (Dyer et al., 2017), and text classification (Balakrishnan et al., 2010). For instance, in sentiment analysis, NLP techniques can be applied to the MD&A section to evaluate the sentiment polarity conveyed by management regarding a company's future prospects. In topic modeling, NLP techniques like Latent Dirichlet Allocation (LDA) can be used to identify the prevalent themes and topics within a report, helping analysts understand the key areas of focus and concern for the company. In text classification, NLP algorithms can be employed to automatically categorize sections of a report or classify the entire document based on specific criteria, such as industry sector or financial performance. In sum, by utilizing NLP techniques, analysts can effectively analyze large numbers of 10-K financial reports in a scalable and efficient manner.

## *Post-Event Volatility as Financial Risk Measurement*

An important financial risk measurement that captures the fluctuations in stock prices following a specific event is post-event volatility (Ito et al., 1998). One common approach for measuring post-event volatility is the naive stock return volatility, defined as the standard deviation of the daily stock returns over a certain period. However, this approach may not adequately capture the effects of specific events on stock price fluctuations.

In the context of post-*event* volatility, the event of primary interest in this paper is the filing of a 10-K financial report. To more accurately measure the impact of such events, we define the post-event return volatility as the root-mean-square error from a Fama and French three-factor model (Fama & French, 1993) for the time period starting from the $6^{th}$ day to the $252^{nd}$ day after the event, with at least 60 daily observations (Loughran & McDonald, 2011; Tsai et al., 2016). By modeling the effect of a report filing on post-event return volatility, we can better understand the relationship between the event and the subsequent changes in stock prices.

The Fama and French three-factor model controls for the market risk premium, size effect, and value effect, providing a way to account for general economic influences that may affect stock returns, thereby isolating the company characteristics. By focusing on the specific fluctuations related to a certain event, such as the filing of a 10-K financial report, we can obtain a more accurate measurement of the volatility that is closely tied to the event itself.

### *Problem Formulation*

Our objective is to harness the Management Discussion and Analysis (MD&A) sections of 10-K financial document as input. Research has highlighted the valuable insights contained within the MD&A section of financial reports. Li (2010) has identified a correlation between the tone of the MD&A and future earnings, while Muslu et al. (2015) have observed that firms whose stock prices do not accurately reflect future earnings are more likely to provide forward-looking information in their MD&A. Additionally, Davis and Tama-Sweet (2012) have found that managers tend to convey more pessimistic information in the MD&A compared to earnings press releases. Collectively, these findings highlight the MD&A as a key resource for understanding a company's financial status. By offering insights into both current conditions and future prospects, it serves as a valuable asset in assessing and potentially predicting the company's financial risk. Each financial document is associated with a post-event volatility value, which indicates the risk level of the corresponding company after the filing of the financial report. We label a 10-K document as “risky” or “non-risky” based on its associated company's post-event volatility value, with “risky” denoting a high value and “non-risky” indicating a low value. The specific threshold values for these labels will be detailed in our Experiment section.

Our aim is to develop a model (FinBERT-XRC) that can accurately predict whether a 10-K document is associated with a risky or non-risky company based on the document's content. To achieve this, we divide our entire dataset of 10-K reports into six distinct sets, each consisting of documents filed in five *training* years and documents filed in a *test* year. The training years include documents from the five years preceding the test year, with test years ranging from 2008 to 2013. This method of partitioning the dataset enhances the robustness of our experiments, as it allows us to evaluate the performance of our FinBERT-XRC model across various time periods.

For each test year, we evaluate the predictive accuracy of our FinBERT-XRC model and compare it with the performance of baseline models. After that, our FinBERT-XRC model generates explanations at the word, sentence, and corpus levels for each test year. These explanations assist users in comprehending the factors that influence the model's predictions, providing insights into the textual elements contributing to the classification of a document as either “risky” or “non-risky”. By providing explanations at different levels of granularity, our FinBERT-XRC model improves the transparency and interpretability of its predictions, enabling users to better grasp the underlying reasons for the model's assessments of company risk. This approach is consistent with the growing emphasis on explainable AI in the financial domain, where comprehending the rationale behind predictions is crucial for informed decision-making.

## Our FinBERT-XRC Model

Figure 1 illustrates the overall structure of our FinBERT-XRC model, which is designed for risk classification in the financial domain while offering explainability for its prediction. We have divided the model into four main components: Sentence Encoder, Sentence-Level Attention Mechanism, Document Encoder, and Attention-Filtered Word Clouds Generator.

### *Sentence Encoder*

#### Text Preprocessing

As depicted in Figure 1a, we start with text preprocessing to prepare our text for input into the FinBERT module. The process includes the following steps.

**Tokenization.** FinBERT uses the BERT tokenizer to convert raw text into tokens, breaking the text into words, subwords, or characters and assigning unique IDs to each.

**Special Tokens.** We add special tokens like `[CLS]` and `[SEP]` to the input. The `[CLS]` token is added at the beginning of the input, while the `[SEP]` token separates different segments, such as paired sentences, and is placed at the end of the input.

**Padding & Truncation.** To ensure a consistent input sequence length, we pad or truncate the sequences as needed. The model requires a fixed input size, so shorter sequences are padded with `[PAD]` tokens, and longer sequences are truncated to fit the model's limit. In our case, the maximum sentence length is 64. Similarly, we set a maximum of 50 sentences per document, as most documents contain fewer than 50 sentences. If a document has more than 50 sentences, we truncate the rest; if it has fewer, we pad it. This approach allows for more efficient utilization of our limited computing resources.

**Attention Mask.** We create an attention mask to differentiate between actual input tokens and padding tokens. This mask is a binary tensor with the same length as the input, where "1" represents real input tokens and "0" represents padding tokens.

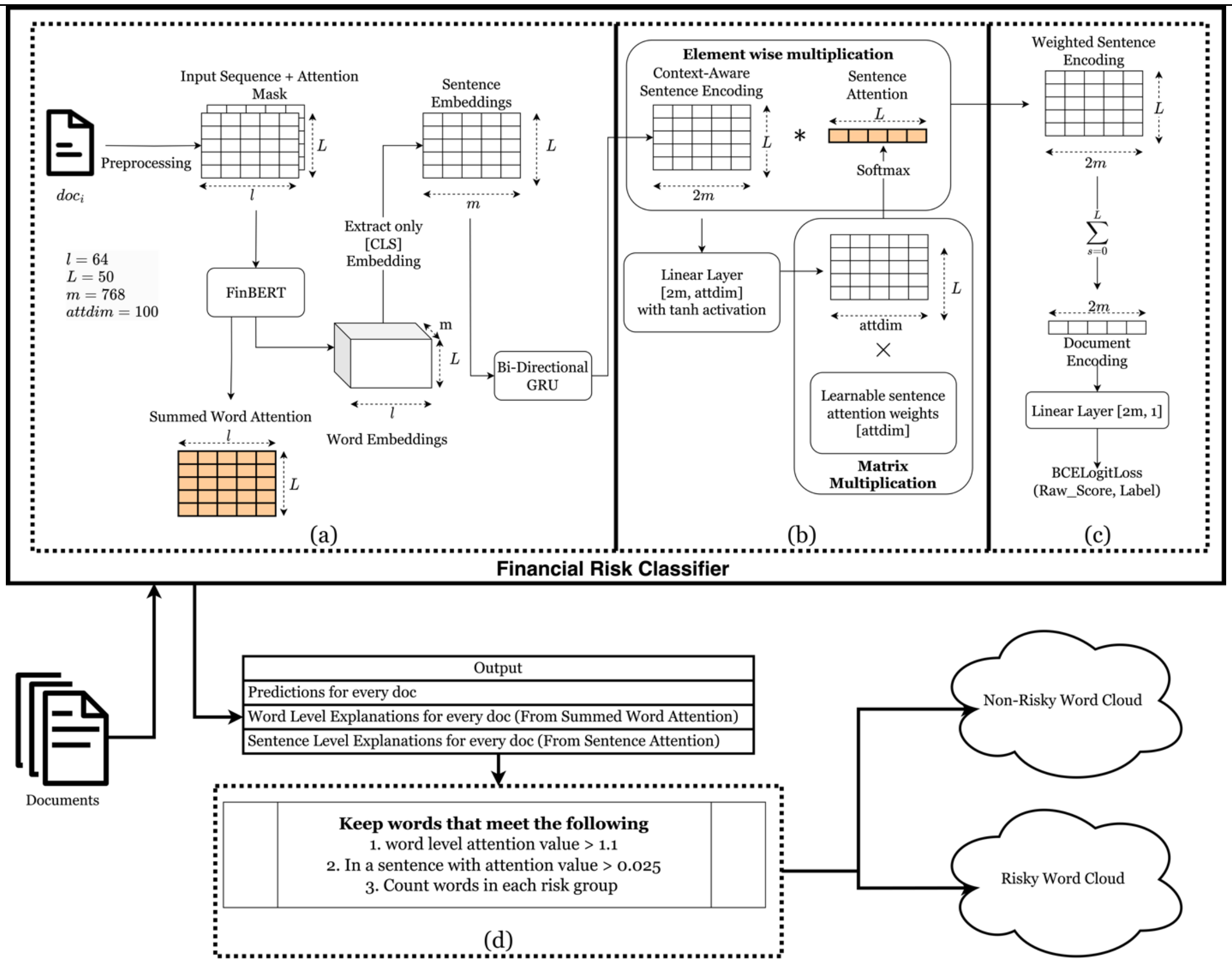

**Figure 1. FinBERT-XRC Architecture**
**(a) Sentence Encoder, (b) Sentence-Level Attention Mechanism,**
**(c) Document Encoder, (d) Attention-Filtered Word Clouds Generator.**

### Sentence Embeddings

*FinBERT* (Liu et al., 2021) is derived from the well-known Bidirectional Encoder Representations from Transformers model (BERT; Devlin et al., 2019), which has been widely used for various natural language processing tasks. FinBERT is trained on a large corpus of financial text, making it better suited for understanding and extracting insights from financial documents, such as 10-K filings, earnings reports, and financial news articles. This specialization allows FinBERT to capture the nuances and intricacies of financial language more effectively than its general-language counterpart, BERT.

As depicted in Figure 1a, we input the prepared data into FinBERT to generate both word embeddings for the text and attention for the words. Word embeddings are continuous vector representations of words that effectively capture their semantic and syntactic properties. These representations enable the model to understand relationships and similarities between words in a given context. On the other hand, attention is a mechanism that assigns importance scores to different words in the text, allowing the model to selectively focus on specific words and their relationships with others. One key aspect of FinBERT is the `[CLS]` token, a special symbol added at the beginning of each input sequence. This token is designed to capture the overall meaning of the input and is often used to represent the entire sequence in tasks like sentiment analysis and classification, hence it is the ideal candidate to represent the embeddings for each sentence. However, to ensure the best performance in our financial risk classification task, we need to fine-tune FinBERT on our dataset. Fine-tuning a pre-trained language model like FinBERT involves training the model on a smaller, domain-specific dataset while employing a lower learning rate. This ensures that the model's internal parameters are only slightly adjusted to learn patterns and relationships relevant to the particular task at hand, without causing drastic changes that may lead to forgetting the general linguistic structure that came from the pre-training on a large corpus of financial text. Specifically, a learning rate of 1e-6 is used for the FinBERT base model, while a rate of 1e-3 is used for the other layers.

The `[CLS]` embeddings, which represent the sentence embeddings, are subsequently passed through a bidirectional Gated Recurrent Unit (GRU; Chung et al., 2014) to capture dependencies and information flow between sentences. This process results in context-aware sentence encoding. At the end of the process, we obtain the context-aware sentence encoding, represented as $h_s$:

$$h_s = \vec{h}_s \oplus \overleftarrow{h}_s = \overrightarrow{\mathrm{GRU}(s_s)} \oplus \overleftarrow{\mathrm{GRU}(s_s)}\,, s = 1, 2, \ldots, L. \quad (1)$$

where $s_s \in \mathbb{R}^m$ is the sentence embeddings from the `[CLS]` token, $h_s \in \mathbb{R}^{2m}$ is the context-aware sentence encoding, and $m = 768$ is the standard embedding size from FinBERT. $L = 50$ is the maximum number of sentences in a document and $s$ is the index of the sentence. $\oplus$ is the concatenation operator. The right arrow (→) and left arrow (←) on top of the operators and variables in the equation represent the forward and backward direction of the GRU, respectively.

The inclusion of the bidirectional GRU allows the model to better understand the relationships between sentences in a given document by capturing both forward and backward dependencies. This context-awareness is important for tasks like risk classification, where the meaning of a sentence is often influenced by the surrounding context. By combining both directions, the model can create a more comprehensive understanding of the relationships between sentences in the document, leading to improved performance in tasks like financial risk classification.

**Summed Word Attention**

The attention output by FinBERT is designed to represent the pairwise attention values between words in the text. However, this representation can be somewhat tricky to understand when trying to determine which specific word is more important than another. To make this information more intuitive and easier to grasp, we consider the idea that a word that receives a lot of attention from other words is more likely to be important. With this in mind, we sum up the attention values along one axis to obtain a single summed attention value for each word in the text. This summed attention value essentially captures the overall importance of a word within the sentence. The summed attention value for each word serves as the foundation for delivering word-level explanations. This approach helps users gain a better understanding of the importance of specific words in the context of financial risk classification, enhancing their ability to interpret the results.

## *Sentence-Level Attention Mechanism*

As seen in Figure 1b, the first step in computing attention scores involves applying a linear transformation to the context-aware sentence encoding, $h_s$. This transformation uses a weight matrix $W \in \mathbb{R}^{a \times 2m}$ and a bias vector $b \in R^a$, where $2m$ denotes the dimension of the context-aware sentence encoding after the bi-directional GRU (in this case, $768 \times 2$) and $a$ represents the attention dimension, which is set to 100. The transformed sentence encoding is then processed by a non-linear activation function, such as the hyperbolic tangent (tanh), ensuring that the value $u_s$ remains within a specific range and introducing non-linearity into the model, as shown in equation 2.

$$u_s = \tanh(W \cdot h_s + b)\,,\ s = 1, 2, \ldots, L. \tag{2}$$

After obtaining the transformed sentence encoding, $u_s$, the next step is to calculate the attention scores. This involves taking the dot product between $u_s \in \mathbb{R}^a$ and a trainable weight vector, $U \in \mathbb{R}^a$ as seen in equation 3. The dot product measures the similarity between the transformed sentence encoding and the weight vector, with higher values indicating greater similarity and hence higher importance of the sentence in the document.

$$\alpha_s = \text{softmax}(u_s^T \cdot U),\ s = 1, 2, \ldots, L. \tag{3}$$

The attention scores are then normalized using the softmax function, converting the scores into a probability distribution. This normalization ensures that the attention scores sum up to 1, allowing them to be used as weights when combining sentence embeddings. The normalized attention score $\alpha_s$ also serves as the basis for the sentence-level explanation for the document. The softmax function is defined in Equation 4.

$$\text{softmax}(x_i) = \frac{\exp(x_i)}{\sum_j \exp(x_j)}. \tag{4}$$

## *Document Encoding*

As shown in Figure 1c, the attention scores from the previous subsection are used to weight the context-aware sentence embeddings. This is achieved by performing element-wise multiplication of the sentence embeddings with their attention scores. This results in a single, fixed-size document representation that captures the most relevant information from the input sentences, based on their attention scores. Subsequently, the weighted sentence embeddings are combined by summing them across the sentence dimension, as shown in Equation 5:

$$d_{\boldsymbol{i}} = \sum_{s=1}^{L} \alpha_s\, h_s, \tag{5}$$

where $d_{\boldsymbol{i}} \in \mathbb{R}^{2m}$ represents the document embeddings for document $\boldsymbol{i}$, $\alpha_s \in \mathbb{R}$ is the attention score for sentence $s$, and $h_s \in \mathbb{R}^{2m}$ is the context-aware embedding for sentence $s$.

Finally, the attended document representation is passed through a fully connected linear layer to generate the final output. During the training phase, this output is compared to the actual target using a loss function called Binary Cross Entropy (BCE) loss, which measures the difference between the predicted and ground-truth values. This loss function is specifically designed for binary classification problems, making it suitable for our use case. When evaluating the model, we do not calculate the loss but instead obtain the prediction by rounding the output. If the output is greater than 0.5, the model classifies the company as risky; if it is less than 0.5, the company is classified as non-risky. This straightforward threshold enables us to categorize companies according to their financial risk, as assessed by our FinBERT-XRC model.

## *Attention-Filtered Word Clouds Generator*

Once the model has been trained, we proceed to evaluate it on the corpus in the test year. We input the test corpus into the Financial Risk Classifier, as depicted in Figure 1. As a result, we obtain the predicted values (risky or not risky), word-level attention values, and sentence-level attention values for *all* the test documents. These values are then fed into the Attention-Filtered Word Clouds Generator, as shown in Figure 1d. Within this generator, we apply a filtering mechanism that retains only the words deemed important in important sentences. This is achieved by considering both the word-level and sentence-level attention values. Subsequently, we count the frequency of each word in the filtered set of important words.

Our approach to generating word clouds is advantageous because it effectively removes irrelevant words from the visualization. This method differs from traditional word cloud generation techniques, which typically rely on purely statistical approaches like TF-IDF. Instead, our model learns the relevant words via the attention layers for the word clouds, leading to a more financial-risk-specific representation.

# Experiments

| Year | Number of Non-Risky Company | Number of Risky Company |
|---|---|---|
| 2003 | 860 | 287 |
| 2004 | 859 | 286 |
| 2005 | 810 | 270 |
| 2006 | 769 | 257 |
| 2007 | 749 | 250 |
| 2008 | 753 | 251 |
| 2009 | 770 | 257 |
| 2010 | 732 | 244 |
| 2011 | 725 | 242 |
| 2012 | 722 | 241 |
| 2013 | 701 | 234 |

**Table 1. Number of Non-Risky and Risky Company for Each Year**

## *Datasets*

In our study, we use the dataset from Tsai (2016). This dataset consists of annual post-event volatility risk calculations for each company, which are then linked to the management's discussion and analysis (MD&A) sections in the corresponding company's 10-K reports.

Our objective is to benchmark our model against the XRR paper (Lin et al., 2021), which also used the same dataset from Tsai (2016). To achieve this, we adopt the same data splitting process as Tsai (2016), who divided the dataset into five bins based on risk levels. Their model is designed to rank companies into different risk levels. However, our focus here is on binary classification to determine whether a company is risky or not. To effectively benchmark our model, we follow the same split as the XRR paper, hence we only consider the top and bottom bins for the classification task. In the context of the XRR model, the ranking task also became a binary classification problem because there are only two classes to rank. For our study, we label companies in the top bin as risky and those in the bottom bin as non-risky. We acknowledge that in real-world applications, it may be unclear whether a company belongs to the top or bottom bin in the unseen data. In such situations, we can simply split the dataset into half, labeling the top 50% as risky and the bottom 50% as non-risky for the training data and evaluate on the unseen data afterwards. We believe that our FinBERT-XRC model will still perform well under such scenarios, as it has already demonstrated superior performance in our experiments compared to the XRR model, the current state of the art. Table 1 presents the number of non-risky and risky companies for each year (with one 10-K document per company). Table 2 provides an overview of the number of training and test data samples (i.e., 10-K documents) where each dataset includes one test year and five training years. The training years include documents from the five years preceding the test year, with test years ranging from 2008 to 2013.

## *Evaluation Metrics*

The evaluation of a model's performance is an important step in any machine learning study. Assessing the effectiveness of a model not only helps us understand its strengths and weaknesses but also enables us to identify areas for improvement and make informed decisions when selecting the most appropriate model for a given task. Furthermore, reliable evaluation metrics allow us to compare different models and approaches objectively, ensuring that we develop robust and accurate solutions for the problems we aim to solve.

Tsai (2016) employs Spearman's Rho (ρ) (Myers et al., 2013) and Kendall's Tau (τ) (Kendall, 1938) as evaluation metrics to assess their model's performance. Both Spearman's Rho and Kendall's Tau are non-parametric rank correlation coefficients, which measure the strength and direction of association between two ranked variables. These metrics are commonly used to evaluate models that perform ranking tasks because they can effectively capture the degree to which two sets of rankings agree with each other, taking into account both the order and the relative differences between the rankings.

| **Dataset** | | **Number of Training Samples** | **Number of Test Samples** |
|---|---|---|---|
| **Training Years** | **Test Year** | | |
| 2003-2007 | 2008 | 5397 | 1004 |
| 2004-2008 | 2009 | 5254 | 1027 |
| 2005-2009 | 2010 | 5136 | 976 |
| 2006-2010 | 2011 | 5032 | 967 |
| 2007-2011 | 2012 | 4973 | 963 |
| 2008-2012 | 2013 | 4937 | 935 |

**Table 2. Dataset Description**

| **Model** | **Evaluation Metrics** | **Test Year** | | | | | |
|---|---|---|---|---|---|---|---|
| | | **2008** | **2009** | **2010** | **2011** | **2012** | **2013** |
| FinBERT-XRC | F1 Score | **0.65** | **0.71** | **0.74** | **0.84** | **0.81** | **0.83** |
| | Kendall's Tau | 0.47 | 0.52 | 0.53 | 0.56 | 0.55 | 0.54 |
| | Spearman's Rho | 0.58 | 0.64 | 0.65 | 0.68 | 0.67 | 0.67 |
| XRR | F1 Score | 0.63 | 0.40 | 0.67 | 0.72 | 0.68 | 0.74 |
| | Kendall's Tau | 0.45 | 0.35 | 0.51 | 0.53 | 0.53 | 0.52 |
| | Spearman's Rho | 0.56 | 0.44 | 0.62 | 0.64 | 0.64 | 0.63 |
| TF-IDF (Logistic Regression) | F1 Score | 0.43 | 0.48 | 0.51 | 0.55 | 0.47 | 0.44 |
| | Kendall's Tau | 0.44 | 0.44 | 0.47 | 0.46 | 0.44 | 0.45 |
| | Spearman's Rho | 0.54 | 0.54 | 0.57 | 0.56 | 0.54 | 0.55 |

**Table 3. F1 Score, Kendall's Tau and Spearman's Rho for Test Year from 2008 to 2013**

However, these metrics are not directly applicable to our case, as we focus binary classification rather than ranking. Nevertheless, we have added the ranking metrics for a more comprehensive evaluation of the various models. It is worth noting that our FinBERT-XRC emerges as the best model based on these metrics as well. In binary classification, accuracy and F1 scores are more commonly used. Accuracy measures the proportion of correct predictions made by the model relative to the total number of predictions, while the F1 score is the harmonic mean of precision and recall. Precision refers to the ratio of true positive predictions to the total number of positive predictions made, whereas recall measures the ratio of true positive predictions to the total number of actual positive instances in the dataset.

F1 score is considered a more suitable evaluation metric for our study, primarily because our dataset is imbalanced. Imbalanced datasets may cause accuracy to be misleading, as the model could achieve high accuracy by merely predicting the majority class. The F1 score, on the other hand, takes both false positives and false negatives into account, providing a more balanced evaluation of the model's performance. By utilizing the F1 score as our primary evaluation metric, we can better assess the effectiveness of our binary classification model in distinguishing between risky and non-risky companies while accounting for any potential class imbalance in the dataset.

## *Baseline Systems and Our FinBERT-XRC Model*

In our study, we benchmark our FinBERT-XRC model against two other models: Term Frequency-Inverse Document Frequency (TF-IDF) based Logistic Regression model and the state-of-the-art XRR model (Lin et al., 2021). These comparisons allow us to evaluate the performance and explainability capabilities of our proposed model in relation to existing methods in the field of financial risk classification.

The TF-IDF (Logistic Regression) model is a traditional machine learning approach that uses the TF-IDF vectorization technique to transform textual data into numerical features, which can then be used as input for a logistic regression classifier. This method quantifies the importance of words in a document by considering both their frequency within the document and their rarity across the entire document collection. By capturing the most distinctive words in the text, the logistic regression classifier is then able to perform risk classification based on the resulting feature vectors.

On the other hand, the XRR model employs a more advanced technique that combines GRUs with *static* word embeddings, such as GLoVe, to process and analyze text data. The use of GRUs enables the model to capture long-term dependencies in the text, while static word embeddings provide a fixed representation of words based on pre-trained contextual information.

Our proposed FinBERT-XRC model utilizes a hybrid approach that combines the strengths of both Transformer and GRU architectures, along with *dynamic* word embeddings that factors in surrounding context when generating word representations. This combination of techniques allows our model to capture richer contextual information and better model the intricate relationships between words in the text.

The XRR model's approach to word-level attention also has limitations, particularly when it comes to generating attention-filtered word clouds. If there are many important words in a sentence, the attention value will be thinly distributed amongst them and be attenuated for each word. This makes it difficult to determine which words are truly important and to establish a static threshold for importance. In contrast, our FinBERT-XRC model employs the attention mechanism from FinBERT, which do not have this issue when there are numerous important words in a sentence.

To ensure the reliability of the results and account for potential randomness, we conducted the experiments 10 times with random initializations, and then calculated the average F1 scores across these repetitions. Table 3 presents the performance of our FinBERT-XRC model and the baseline models in terms of F1 Score. The best F1 score for each test year is highlighted in bold, and it is evident that our FinBERT-XRC model consistently outperforms the other two models from 2008 to 2013. Notably, the XRR model, which is considered the state-of-the-art in classifying financial risk documents, performs significantly worse in comparison to our FinBERT-XRC model. According to Table 3, the XRR model lags behind our FinBERT-XRC model by an average of 12.3% in terms of F1 score. As expected, the TF-IDF (Logistic Regression) model exhibits the largest performance gap, lagging behind by 28% in terms of F1 score.

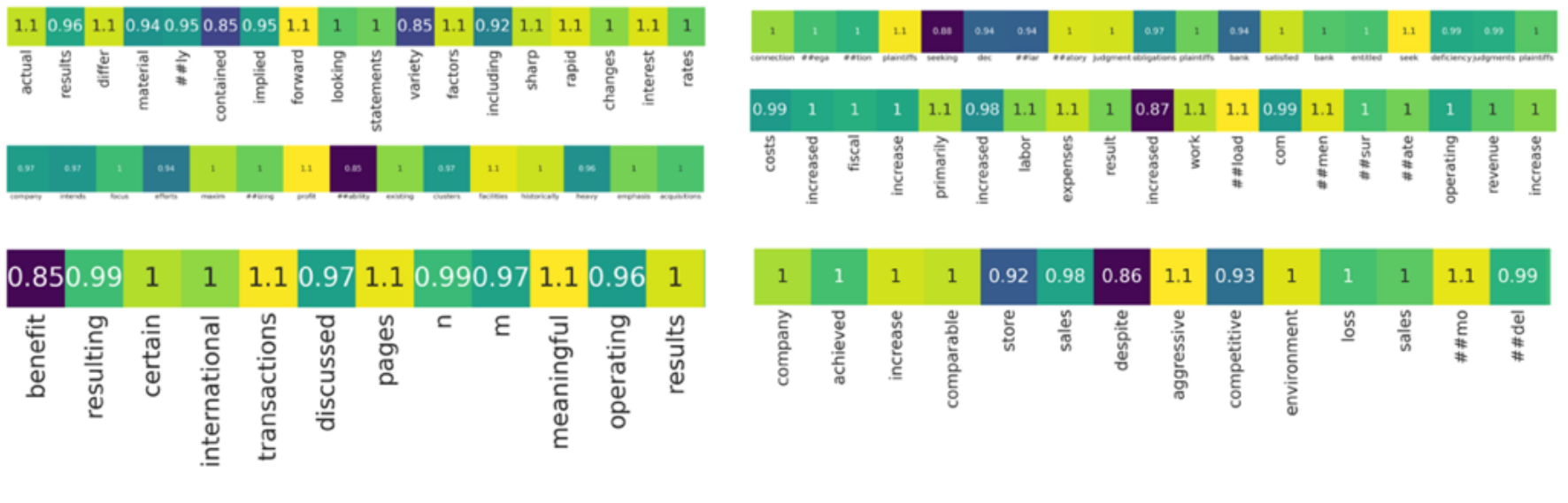

(a) (b)

**Figure 2. Word Level Explanation Demonstration for (a) Non-Risky Documents, (b) Risky Documents**

## *Explanation by Our FinBERT-XRC Model*

In this section, we will delve into the explainability of the model. Our FinBERT-XRC model provides explanations at three different levels: word, sentence, and corpus levels.

### Word Level Explanation

We have discussed the calculation of the summed word attention from the FinBERT attention output which serves as the basis for our word-level explanation. Figure 2 demonstrates the word-level explanations provided by our FinBERT-XRC model. In the Non-Risky group as shown in Figure 2a, words like "forward", "meaningful", and "profit" exhibit high attention values. On the other hand, for the Risky group as shown in Figure 2b, words such as "aggressive", "plaintiffs", and "primarily increased labor expenses" display high attention values. As demonstrated, our FinBERT-XRC model effectively reveals meaningful insights at the word level by identifying words that contribute to the risk attributes, thereby enhancing our understanding of the factors driving financial risk.

### Sentence Level Explanation

| Rank | Attention Value | Sentence |
|---|---|---|
| 1 | 0.05715 | fiscal compared to fiscal in fiscal we experienced a **net loss of million** on revenues of million as compared to a net income of million on revenues of million for fiscal this represents a **decrease in revenues** of. |
| 2 | 0.05715 | revenues are in the form of fees which are earned under contracts with mri facilities and physical rehabilitation practices |
| 3 | 0.05715 | this was due **mostly to decreased product sales and management fees**. |
| 4 | 0.05715 | our consolidated operating results decreased by million to an **operating loss of million** for fiscal as compared to an operating income of million for fiscal discussion of operating results of medical equipment segment fiscal compared to fiscal revenues attributable to **our medical equipment segment decreased by to million in fiscal** from million in fiscal reflecting a decrease in product sales revenues of from million in fiscal to million in fiscal offset by an increase in service revenue of from million in fiscal to million in fiscal this decline in revenues was attributable to a **reduction in sales** of our upright tm mri. |
| 5 | 0.05714 | hmca commenced operations in july and generates revenues from providing comprehensive management services including development administration accounting billing and collection services together with office space medical equipment supplies and non medical personnel to its clients. |

**Table 4. Example of Sentence Level Explanation for a Risky Company**

Table 4 presents the top five sentences with the highest attention values from a risky company. We have highlighted some of the keywords in bold to emphasize the relevance of each sentence. By referring to Table 4, we can readily discern that the company is experiencing financial difficulties due to a decrease in product sales. This succinct summary offers a glimpse into a company's financial status, showcasing the value of pinpointing significant sentences for better understanding and decision-making.

### Corpus Level Explanation

In this paper, one of our primary objectives is to showcase explainability at the corpus level. Empirically, we determined that a cutoff for word attention at 1.1 and for sentence attention at 0.025 results in the most meaningful word clouds. However, do note that these thresholds can be subjective. Figure 3 demonstrates the risky word cloud for the year 2001, along with a real-life case study of a company called *AMCON*. As seen in Figure 3a, the word "AMCON" stands out prominently in the word cloud, indicating that it carries significant risk elements. Figure 3b displays the stock price for AMCON from year 2000 to year 2001, showing its value dropping from $60 to $20, hence substantiates the risk implied by the word cloud. This demonstration highlights the value of providing corpus-level explanations, as it enables the identification of overarching themes and trends within the corpus. In this instance, it even pinpoints a specific company that exhibits risky characteristics, emphasizing the practical utility of incorporating such explanations in financial risk analysis.

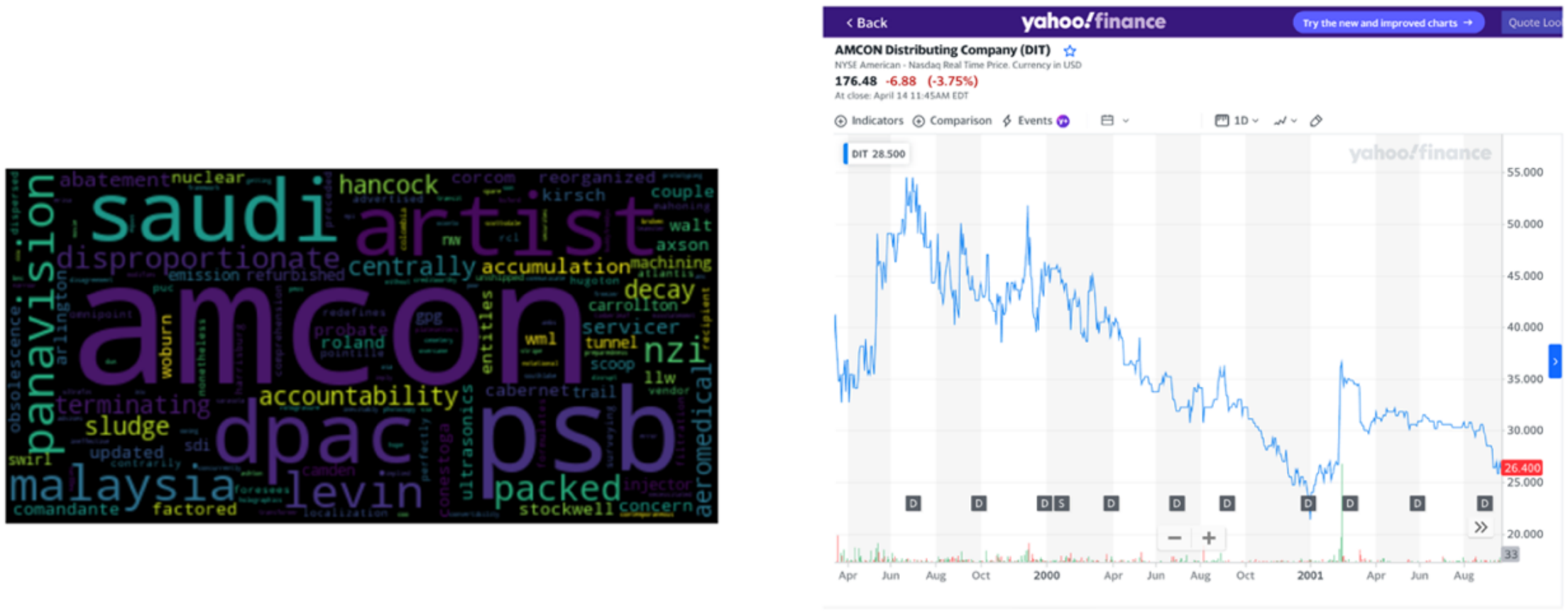

(a) (b)

**Figure 3. Example of Risky Word Cloud**

**(a) Risky Word Cloud for Year 2001, (b) AMCON Stock Price from 2000 to 2001**

## *Discussion of Results*

Our experimental results have demonstrated the superior performance of our FinBERT-XRC model in the financial risk classification task when compared to both the XRR and the TF-IDF (Logistic Regression) models. In this section, we discuss the results by comparing our model to these two approaches.

First, let's consider the comparison between our FinBERT-XRC model and the TF-IDF (Logistic Regression) model. The TF-IDF (Logistic Regression) model is a more traditional approach, relying on term frequency-inverse document frequency (TF-IDF) to weigh the importance of words in documents and then using logistic regression for classification. While this approach has been widely used in text classification tasks, it does not take into account the semantic relationships between words and the overall context in which they appear. In contrast, our FinBERT-XRC model leverages FinBERT, which is pre-trained on a large corpus of financial documents, to generate context-aware word embeddings. By using the `[CLS]` token to represent sentence embeddings, our model can better capture the overall meaning of the text, leading to improved classification performance.

Next, we compare our FinBERT-XRC model to the XRR model, which utilizes fine-tuned GLoVe embeddings. Although the XRR model employs pre-trained word embeddings, they are static and do not account for the surrounding context. As a result, the XRR model may struggle to capture the nuances in financial documents that our FinBERT-XRC model can handle more effectively. Furthermore, our FinBERT-XRC model benefits from the combination of Transformer for long range dependencies and GRU to handle sequential information, which allows it to better understand and represent the hierarchical structure of the text. This hybrid approach enables our model to outperform the XRR model, which relies solely on GRU.

Lastly, we turn our focus to the importance of explanations. The explanations provided by our FinBERT-XRC have the potential to reshape how users interpret information and understand the underlying data (Bauer et al., 2023), particularly in contexts like investment where AI-driven predictions are integral to making informed investment decisions (Ban et al., 2018). Explanations by the model help to create a channel to investigate how did the model reached a certain prediction. Furthermore, in regulatory contexts where fairness is prioritized over accuracy (Fu et al., 2022), these explanations can empower end-users to modify the model's decision without undermining its overall performance, should the explanation appear unsound. In institutions such as banks, where financial authorities require credit scoring models to be explainable (De Lange et al., 2022), our FinBERT-XRC model could be integrated as a supplementary component. It can provide additional insights for their credit rating system while aligning with the necessary regulations since the model itself is already interpretable. In the field of Information Systems (IS), the interaction between humans and AI-enabled systems is emerging as a growing trend (Rzepka & Berger, 2018). Unlike a black-box model, which lacks means of interaction, providing interpretability in an AI system enables end-users to engage with the underlying algorithms and gain a deeper understanding of them. This, in turn, opens up new opportunities for IS researchers to explore human behavior, cognition, feedback loops between machines and humans, and decision-making processes.

## Conclusion and Future Work

In conclusion, our FinBERT-XRC model addresses challenges in the field of finance, particularly in explainable AI, by offering a powerful and interpretable approach to financial risk classification. It not only outperforms the existing top-tier approach but also imparts insightful explanations on various textual levels. Financial professionals can leverage this information to craft more strategic investment decisions, analysts can gain a deeper understanding of market trends and underlying risks, and even individual investors can benefit by obtaining clearer and more actionable insights into their investment portfolios. As a future endeavor, we plan to explore methods to evaluate our model's explainability. We recognize the inherent challenge in objectively assessing such explanations, which often can be subjective. This remains an open research question to this day. One possible solution is to engage domain experts to create "gold labels" for evaluating our model's explanations objectively. Additionally, we are keen to refine our risk measurement approach. Our current use of volatility as a risk metric may not fully capture the true nature of financial risk, as it inappropriately classifies upside risk as "risk" when it leads to profits. Our focus should instead be on downside risk, which could lead to significant financial loss for clients. We will employ skewness, kurtosis, and Value-at-Risk (VaR) in our analysis. Left skewness highlights the potential for negative returns, while kurtosis emphasizes the risk of extreme negative returns. VaR quantifies the potential loss in the value of a portfolio, pinpointing the magnitude of downside risk. Together, these metrics would form a more nuanced understanding of financial risk, enhancing the utility and robustness of our model.